LLUMS QUE NO SÓN NOMÉS LLUMS

# Arriba la comunicació per llum visible

**Les comunicacions per llum visible o 'Visible Light Communications' (VLC) són un nou paradigma en comunicacions sense fils. Les característiques que presenta aquesta tecnologia, que utilitza els dispositius d'il·luminació basats en díodes emissors de llum com a elements transmissors, fa que es pugui considerar un complement dels actuals sistemes de comunicació inalàmbrics.**

En aquest article es resumeixen les principals característiques de la tecnologia VLC, els seus principals camps d'aplicació i les activitats dutes a terme al camp de l'estandardització. Finalment, es presenta una descripció del demostrador SILENCE ('Software deflned Light Communication systEm') que ha realitzat el Centre Tecnològic de Telecomunicacions de Catalunya (CTTC) i les seves prestacions. Aquest sistema és la primera implementació real de l'estàndard IEEE 802.15.7 per a comunicacions VLC.

Les noves tendències en sistemes d'il·luminació arreu del món estan basades en l'ús de díodes emissors de llum, més popularment coneguts com llums LED. L'evolució que han experimentat aquests dispositius als darrers anys ha permès el desenvolupament de nous elements d'il·luminació a un cost competitiu. A més, la seva eficiència energètica i llarga vida útil constitueixen una alternativa seriosa per al recanvi de les tradicionals bombetes i llums fluorescents. En aquest sentit, la porció de mercat dels llums LED augmenta any rere any i es preveu que la majoria dels sistemes d'il·luminació utilitzin aquest tipus de dispositius l'any 2018 [1].

Tanmateix, les propietats dels LED esmentades abans han fet que es trobin usos alternatius per a aquests dispositius. En l'última dècada l'interès de la comunitat científica pels LEDs ha crescut degut al fet que es poden utilitzar com a elements transmissors d'informació sense perdre

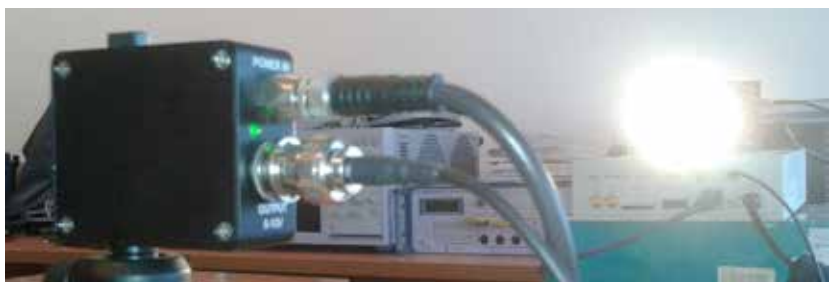

↑ Demostrador SILENCE.

la seva principal funcionalitat de font d'il·luminació [2]. Aquest fet ha contribuït a l'aparició d'un nou paradigma en les comunicacions sense fils: les comunicacions per llum visible, o més conegudes com 'Visible Light Communications'. En les comunicacions VLC la informació a transmetre serveix per modular la senyal d'alimentació dels llums LED, de manera que s'utilitza l'ona de llum com a senyal portador. És a través de la variació d'intensitat d'aquesta ona com es transmet la informació. En la recepció s'utilitza un foto-detector per extreure'n la informació del senyal òptic. També es poden utilitzar com a elements receptors, sensors CMOS o CCD, dispositius que s'utilitzen en les càmeres digitals, com ara els que porten els telèfons mòbils.

## Característiques de les comunicacions VLC

Les comunicacions VLC presenten nombrosos avantatges respecte als sistemes tradicionals de radiofreqüència. Entre d'altres podem destacar:

- Disponibilitat d'una gran amplada de banda a una part de l'espectre, el de la llum visible, sense llicenciar.
- No causa interferències amb els actuals sistemes de comunicació basats en RF, de manera que es pot utilitzar en entorns sensibles com hospitals, avions o fàbriques.
- Capacitat per oferir un servei de comunicació a un cost reduït degut al futur desplegament massiu de sistemes d'il·luminació LED.
- L'ús de la tecnologia VLC no comporta cap preocupació per a la salut ja que usa llum visible.
- Reduïda empremta ecològica derivada del seu baix consum energètic.
- En condicions normals d'il·luminació [3] s'aconsegueix prou intensitat de senyal com per establir la comunicació, de manera que no es requereix l'ús de dispositius especials d'il·luminació que poden elevar els costos de desplegament.

Hi ha altres propietats de les comunicacions VLC que tot i constituir a priori un desavantatge davant les comunicacions de radiofreqüència, en segons quines condicions es transformen en avantatges. El fet que la llum estigui confinada en l'espai on es realitza la transmissió i no pugui travessar ni parets ni envans fa que el seu abast sigui molt localitzat. Tanmateix, aquest comportament pot resultar un avantatge quan es parla de comunicacions interiors ja que s'aconsegueix un alt grau de privacitat. Terceres persones no poden interceptar la comunicació sense estar presents físicament al lloc on es produeix la comunicació. Addicionalment, s'aconsegueix una millor reutilització espacial dels recursos freqüencials, de manera que la interferència entre usuaris està més controlada. Actualment, el desplegament massiu de punts d'accés de xarxes Wi-Fi provoca la col·lisió entre usuaris, de manera que la seva amplada de banda efectiva es veu reduïda.

Tot i així, un dels problemes que aquesta jove tecnologia presenta i que ha de solucionar durant el seu procés d'evolució és la possibilitat de poder ser interferida per altres fonts de llum de gran intensitat que poden saturar el receptor. En aquestes condicions el foto-detector no és capaç de processar correctament els senyals de llum i extreure'n la informació útil. Aquesta circumstància és menys probable que ocorri en entorns interiors, perquè els sistemes d'il·luminació estan dissenyats per tal de proveir un nivell d'il·luminació que no resulta tan extrem per al funcionament adequat de l'element receptor.

D'altra banda, es discuteix aferrissadament sobre el desenvolupament del canal de retorn. Per tal de fer possible la bidireccionalitat de les comunicacions VLC, actualment es consideren dues vessants. Per una banda, hi ha propostes per dotar el receptor amb un transmissor LED, de forma que pugui comunicar-se a través d'un enllaç VLC. Per l'altra, es parla d'incorporar la tecnologia d'infraroig. Tot i així, ambdues solucions poden divergir segons l'escenari on siguin necessàries. Per exemple, en entorns d'interiors, com oficines, l'opció d'utilitzar un canal de retorn amb infraroig podria resultar més adient perquè la comunicació seria no visible i l'usuari no percebria noves fonts de llum a l'entorn que podrien resultar molestes. En canvi, en un entorn de comunicacions subaquàtiques entre submarinistes, l'existència d'altres fonts de llum no és tan molesta, fent així adequat l'ús d'un sistema complet utilitzant VLC en ambdues direccions. En conclusió, la creació del canal de retorn és un tema que no està tancat i per tant cal fomentar la discussió sobre la seva implementació.

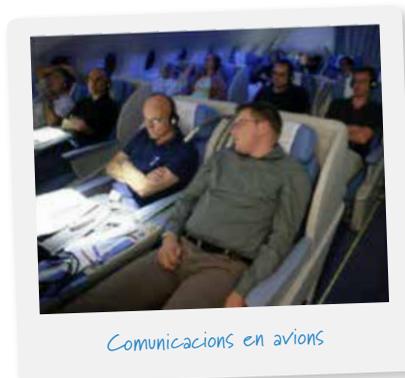
Comunicacions en avions

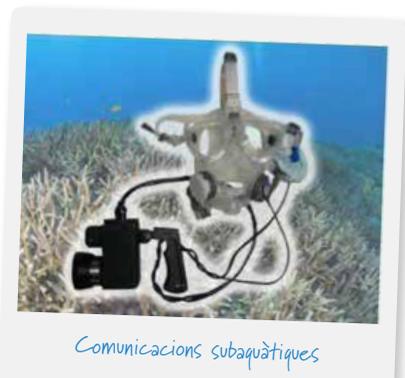
Comunicacions subaquàtiques

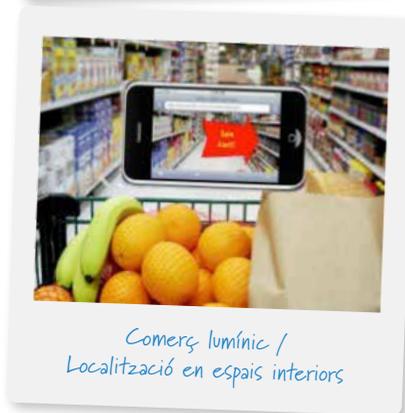
Comerç lumínic / Localització en espais interiors

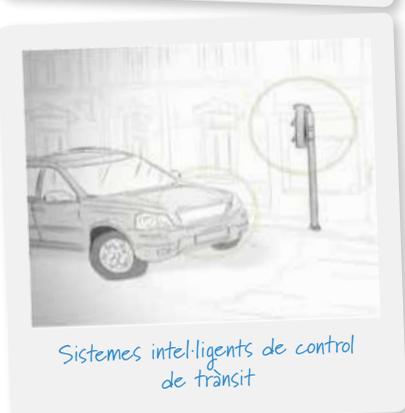
Sistemes intel·ligents de control de trànsit

**Figura 1.** Exemples d'aplicacions de la tecnologia VLC. Imatges extretes de [4].

## Camps d'aplicació

L'estat inicial de desenvolupament de la tecnologia VLC unit amb la necessitat de trobar el seu espai i les sinèrgies amb les altres tecnologies que habiten l'ecosistema digital han fet que s'identifiqui una gran varietat d'àrees d'aplicació [4] (**Figura 1**). Es preveu que en un futur molt pròxim apareguin al mercat solucions comercials per als escenaris plantejats. Aquests escenaris es podrien classificar segons la seva necessitat de velocitat de transferència. S'identifiquen dos camps d'aplicació amb una alta necessitat de velocitat:

- **Comunicacions en espais interiors per dotar d'accés a Internet o continguts multimèdia.** La tecnologia VLC pot oferir una alternativa a la problemàtica de la sobreocupació de l'espectre i les seves capacitats de reutilització en entorns on hi ha una alta concentració de comunicacions i tecnologies, com ara els habitatges o els entorns d'oficines. A més, VLC és una alternativa interessant per a les comunicacions sense fils en entorns on la radiofreqüència no és adient, com ara l'interior d'avions o en hospitals.

- **Comunicacions subaquàtiques.** Alguns dels requisits més importants a l'hora de desenvolupar una tecnologia de comunicació subaquàtica viable són la transmissió fiable de grans volums de dades, a més de la possibilitat de controlar remotament robots i vehicles sota l'aigua. En aquest sentit, VLC pot ser una alternativa molt interessant per solucionar la falta de velocitat que presenta l'ús d'ones acústiques i els problemes d'abast que presenten les solucions basades en radiofreqüència.

Pel que fa als camps d'aplicació amb unes necessitats més reduïdes de velocitats, es poden identificar els següents:

- **Sistemes intel·ligents de control del trànsit.** Els semàfors poden transmetre informació sobre l'estat del trànsit a cotxes o trens de forma remota. En el cas del trànsit urbà aquesta informació es pot combinar amb informació GPS amb l'objectiu d'informar als conductors i evitar problemes de congestió, amb la conseqüent reducció del consum de combustible i emissions de $CO_2$.

- **Comerç lumínic.** Els llums dels anuncis que hi ha al carrer, dels cartells o la pròpia il·luminació dels establiments poden ser utilitzats per enviar informació comercial al client. Aquesta informació podria consistir en possibles ofertes per-

21

sonals, nous serveis i productes o descomptes addicionals que millorarien la satisfacció del client i revertir als resultats comercials dels establiments.

- **Sistemes de localització en espais interiors.** Aquesta aplicació suposa una solució al problema que experimenta el GPS en entorns interiors, on perd la seva cobertura. Els llums dels edificis poden transmetre un identificador que serveixi per localitzar la posició de l'individu dins de l'edifici. Aquest sistema es pot combinar amb l'anterior aplicació de comerç lumínic per guiar als clients dins dels comerços i oferir ofertes en funció de la localització. Paral·lelament, els responsables dels establiments podrien obtenir estadístiques per veure les preferències dels clients i millorar les seves estratègies de màrqueting i, addicionalment, en edificis públics, per indicar on es pot realitzar un tràmit determinat.

### Activitats normalitzadores

Com s'ha comentat anteriorment, el potencial de desenvolupament que presenta el mercat de les llums LED suposa una oportunitat per a les comunicacions VLC. Aquesta oportunitat ha estat reconeguda per rellevants comitès d'estandardització.

Al Japó, el 'Visible Light Communications Consortium' (VLCC) [5] treballa des del 2003 en l'estandardització d'aquest tipus de comunicacions. Entre els membres d'aquest consorci es troben importants empreses com Samsung, Casio, Panasonic, NTT Docomo així com grups de recerca de diferents universitats japoneses (Keio University, Waseda University and Tokyo University). L'any 2007 va presentar les normes CP-1221 i CP-1222 a l'associació japonesa d'indústries electròniques i de les tecnologies de la informació (JEITA). Al 2008, el VLCC va formar una aliança amb l'associació de tecnologia infraroja anomenada IrDA. D'aquesta unió va sorgir una nova especificació el 2009 que adaptava la normativa IrDA per fer servir dispositius compatibles amb l'especificació IrDA com a sistema VLC.

Les activitats normalitzadores del 'Institute of Electrical and Electronics Engineers' (IEEE) van començar l'any 2009 [6]. El grup de treball IEEE 802.15.7 va treballar per a la definició d'un nou estàndard de comunicacions VLC de capa física i d'enllaç que va ser aprovat al 2011. Aquest grup de treball ha considerat per a l'estàndard les contribucions d'empreses com Intel, Samsung i Siemens i dels grups de recerca que van participar al projecte OMEGA [7], finançat per la Unió Europea.

A l'octubre de 2011 es va formar el Li-Fi Consortium [8] —Li-Fi significa 'Light Fidelity', similar al concepte Wi-Fi—. L'objectiu d'aquest consorci és promoure el desenvolupament de la tecnologia VLC. Això ho aconsegueixen convidant experts en la tecnologia, fabricants i participants als grups d'estandarització a fòrums periòdics de discussió per trobar solucions als reptes i necessitats d'aquesta tecnologia emergent. Tanmateix, s'espera més moviment en aquest àmbit ja que cap d'aquestes especificacions ha comptat amb la col·laboració activa dels fabricants de LED. De fet, aquesta sinèrgia sembla necessària ja que algunes de les empreses que comencen a comercialitzar productes que utilitzen solucions basades en tecnologia VLC, com ara LVXSystem [9] i ByteLight [10], tenen acords subscrits amb fabricants de llums LED.

### El demostrador SILENCE

El potencial que presenta aquesta tecnologia dins del context de les ciutats intel·ligents, unit a la seva component d'innovació i repte tecnològic ha fet despertar l'interès del CTTC, que treballa des del 2012 en aquest camp. Un exemple dels resultats del seu treball es el demostrador SILENCE, el primer sistema real que utilitza la tecnologia VLC basada en l'estàndard IEEE 802.15.7 [11].

Per a la construcció d'aquest demostrador s'ha utilitzat el concepte 'Software Defined Radio' (SDR), on les funcions de processament del senyal corresponen a un processador de propòsit general (PPG), mentre que les funcions de RF i de conversió del senyal analògic-digital, digital-analògic (A/D, D/A) van a càrrec d'un maquinari programable.

Gràcies a la utilització de les tècniques SDR, SILENCE està dotat d'un alt grau de flexibilitat i modularitat respecte a les tradicionals implementacions en maquinari utilitzant circuits dedicats. Aquesta característica li permet evolucionar i acollir les millores futures sense requerir un llarg temps de desenvolupament, constituint així una plataforma de desenvolupament per provar diferents esquemes i protocols de comunicació utilitzant diferents algoritmes de processament del senyal.

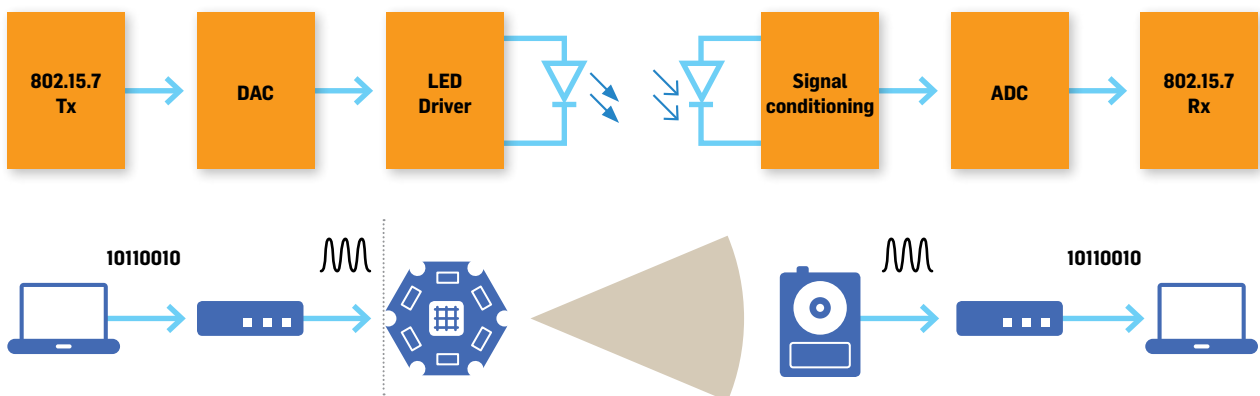

**Figura 2.** Diagrama de blocs del demostrador SILENCE.

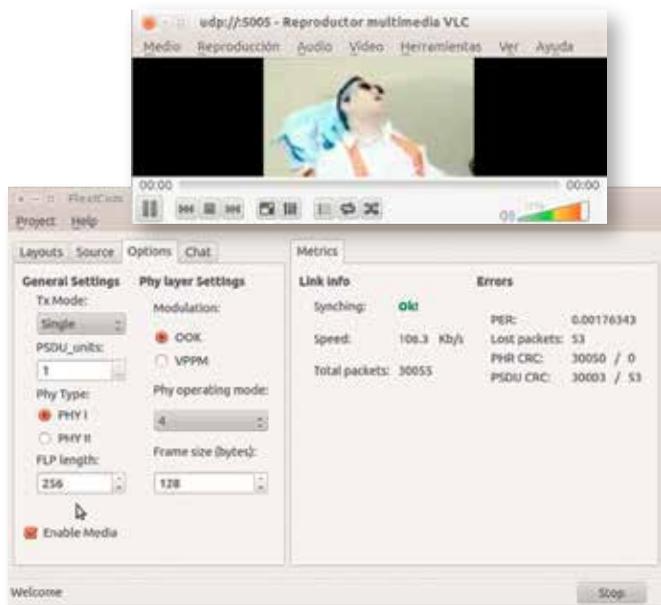

**Figura 3.** Exemple de recepció de vídeo amb el demostrador SILENCE. Detall de la interfície gràfica d'usuari que configura el sistema.

SILENCE consta de dos dispositius, l'un transmissor i l'altre receptor, capaços d'establir una comunicació unidireccional en temps real utilitzant qualsevol mode de funcionament especificat per la capa física I (PHY I) definida en l'estàndard IEEE 802.15.7. Cal mencionar que la part de programari del sistema també està preparada per poder fer una transmissió d'acord a l'especificació de capa física II (PHY II) de l'estàndard. Però el LED utilitzat limita les prestacions del sistema especialment en amplada de banda, fent possible només la transmissió amb els modes definits per PHY I. En la banda transmissora, un ordinador construeix la trama d'informació tal i com s'especifica a l'estàndard i dóna les mostres a la 'Universal Series Radio Peripheral' (USRP), que és el maquinari programable que genera el senyal modulat utilitzat per controlar el LED. En la part receptora, un fotodíode PIN recull la llum procedent del LED i la converteix en un corrent elèctric, que després de passar per un procés d'amplificació, es mostreja a l'USRP receptora, passant del domini analògic al digital. L'USRP envia les mostres a l'ordinador receptor i es procedeix a realitzar el processat de les dades rebudes per extreure'n la informació continguda (**Figura 2**).

Actualment SILENCE està dotat de capacitats multimèdia i es pot configurar per fer una transmissió de text a través d'un xat o de vídeo de baixa resolució en temps real ('streaming'), des del dispositiu transmissor a diversos receptors. Les prestacions del sistema varien en funció del mode de funcionament, és a dir, si es transmet text o vídeo. A diferència del text, la reproducció del vídeo en temps real fa servir un software de visualització de vídeo aliè al sistema. Amb SILENCE s'han aconseguit realitzar transmissions de text a distàncies de 8 m i de vídeo a 1,5 m amb una taxa d'error de paquet del 0,1% amb visió directa entre el transmissor i el receptor. La màxima velocitat de transmissió d'informació útil, és a dir, bits d'informació sense redundància i capçaleres, és superior als 100 Kbps (**Figura 3**).

Els sistemes de comunicacions VLC es troben en una fase embrionària i encara hi ha moltes qüestions per formular i solucionar entorn d'aquesta tecnologia emergent. El previsible desenvolupament dels sistemes basats en llums LED propiciarà que aquestes preguntes trobin resposta ja que representa una oportunitat excitant per a diferents actuadors del món empresarial, des de la indústria lumínica fins a les empreses de telecomunicacions i els creadors d'aplicacions mòbils. En aquest sentit, les característiques de la plataforma SILENCE, en termes de flexibilitat i modularitat, fan que el CTTC disposi d'una eina molt útil per al desenvolupament d'activitats de recerca al camp de VLC.

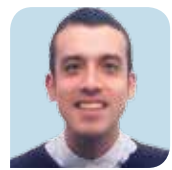

**Jorge Baranda**
Enginyer de telecomunicació
Enginyer de recerca del CTTC

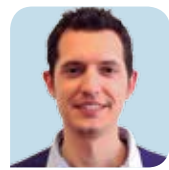

**Pol Henarejos**
Enginyer de telecomunicació
Enginyer de recerca del CTTC

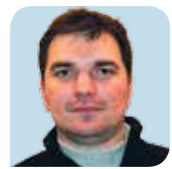

**Ciprian George Gavrincea**
Doctor enginyer de telecomunicació i investigador associat de l'Àrea de Enginyeria del CTTC

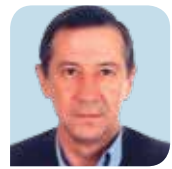

**Miguel Angel Lagunas**
Director del CTTC
Col·legiat núm. 579